\documentclass[letterpaper, 10 pt, conference]{ieeeconf}  

\IEEEoverridecommandlockouts                              
\usepackage{graphics} 
\usepackage{epsfig} 
\usepackage{mathptmx} 
\usepackage{times} 
\usepackage{amssymb}  
\usepackage{mathtools}
\usepackage{dblfloatfix}
\usepackage{color}
\usepackage{booktabs,tabularx}
\usepackage{optidef}
\usepackage[font=small,skip=3pt]{caption}
\usepackage{color,soul}

\usepackage{siunitx}
\usepackage{cite}
\usepackage{url}

\usepackage{subfigure}
\renewcommand{\thefigure}{\arabic{figure}}
\title{\LARGE \bf
Low-Cost 3D printed, Biocompatible Ionic Polymer Membranes for Soft Actuators
}

\author{Nils Trümpler$^{1,\dag}$, Ryo Kanno$^{2,3,\dag}$, \textit{Student Member, IEEE}, Niu David$^{1}$, Anja Huch$^{4}$, Pham Huy Nguyen$^{2,3,5}$, \\ \textit{Member, IEEE}, Maksims Jurinovs$^{6}$, Gustav Nyström$^{1,4}$, \textit{Member, IEEE}, Sergejs Gaidukovs$^{6}$, \\ Mirko Kovac$^{2,3,5,*}$, \textit{Member, IEEE}
\thanks{$^{1}$N. Trümpler, D. Niu, and N. Gustav are with the Eidgenössische Technische Hochschule Zürich, 8092 Zurich, Switzerland.}
\thanks{$^{2}$R. Kanno, P. H. Nguyen, and M. Kovac are with the Laboratory of Sustainability Robotics, Empa - Swiss Federal Laboratories for Materials Science and Technology, 8600 Dübendorf, Switzerland. {\tt\small $\{$mirko.kovac$\}$@empa.ch}}
\thanks{$^{3}$R. Kanno, P. H. Nguyen, and M. Kovac are with the École Polytechnique Fédérale de Lausanne, 1005 Lausanne, Switzerland.}
\thanks{$^{4}$A. Huch and G. Nyström are with the Cellulose \& Wood Materials Laboratory, Empa - Swiss Federal Laboratories for Materials Science and Technology, 8600 Duebendorf, Switzerland.}
\thanks{$^{5}$P. H. Nguyen and M. Kovac are with the Aerial Robotics Laboratory, Imperial College London, South Kensington Campus, London, SW7 2AZ, United Kingdom.}
\thanks{$^{6}$M. Jurinovs and S. Gaidukovs are with Institute of Chemistry and Chemical Technology, Faculty of Natural Sciences and Technology, Riga Technical University, P. Valdena Str. 3, LV-1048, Riga, Latvia}
\thanks{$*$ Address correspondences to these authors. $\dag$ These authors contributed equally.}
}
\begin{document}

\maketitle
\thispagestyle{empty}
\pagestyle{empty}

\begin{abstract}
Ionic polymer actuators, in essence, consist of ion exchange polymers sandwiched between layers of electrodes. They have recently gained recognition as promising candidates for soft actuators due to their lightweight nature, noise-free operation, and low-driving voltages. However, the materials traditionally utilized to develop them are often not human/environmentally friendly. Thus, to address this issue, researchers have been focusing on developing biocompatible versions of this actuator. Despite this, such actuators still face challenges in achieving high performance, in payload capacity, bending capabilities, and response time. In this paper, we present a biocompatible ionic polymer actuator whose membrane is fully 3D printed utilizing a direct ink writing method. The structure of the printed membranes consists of biodegradable ionic fluid encapsulated within layers of activated carbon polymers. From the microscopic observations of its structure, we confirmed that the ionic polymer is well encapsulated. The actuators can achieve a bending performance of up to 124$^\circ$ (curvature of 0.82 $\text{cm}^{-1}$), which, to our knowledge, is the highest curvature attained by any bending ionic polymer actuator to date. It can operate comfortably up to a 2 Hz driving frequency and can achieve blocked forces of up to 0.76 mN. Our results showcase a promising, high-performing biocompatible ionic polymer actuator, whose membrane can be easily manufactured in a single step using a standard FDM 3D printer. This approach paves the way for creating customized designs for functional soft robotic applications, including human-interactive devices, in the near future.

\end{abstract}
\section{INTRODUCTION}
\begin{figure}[t]
\centering
\includegraphics[width=0.49\textwidth]{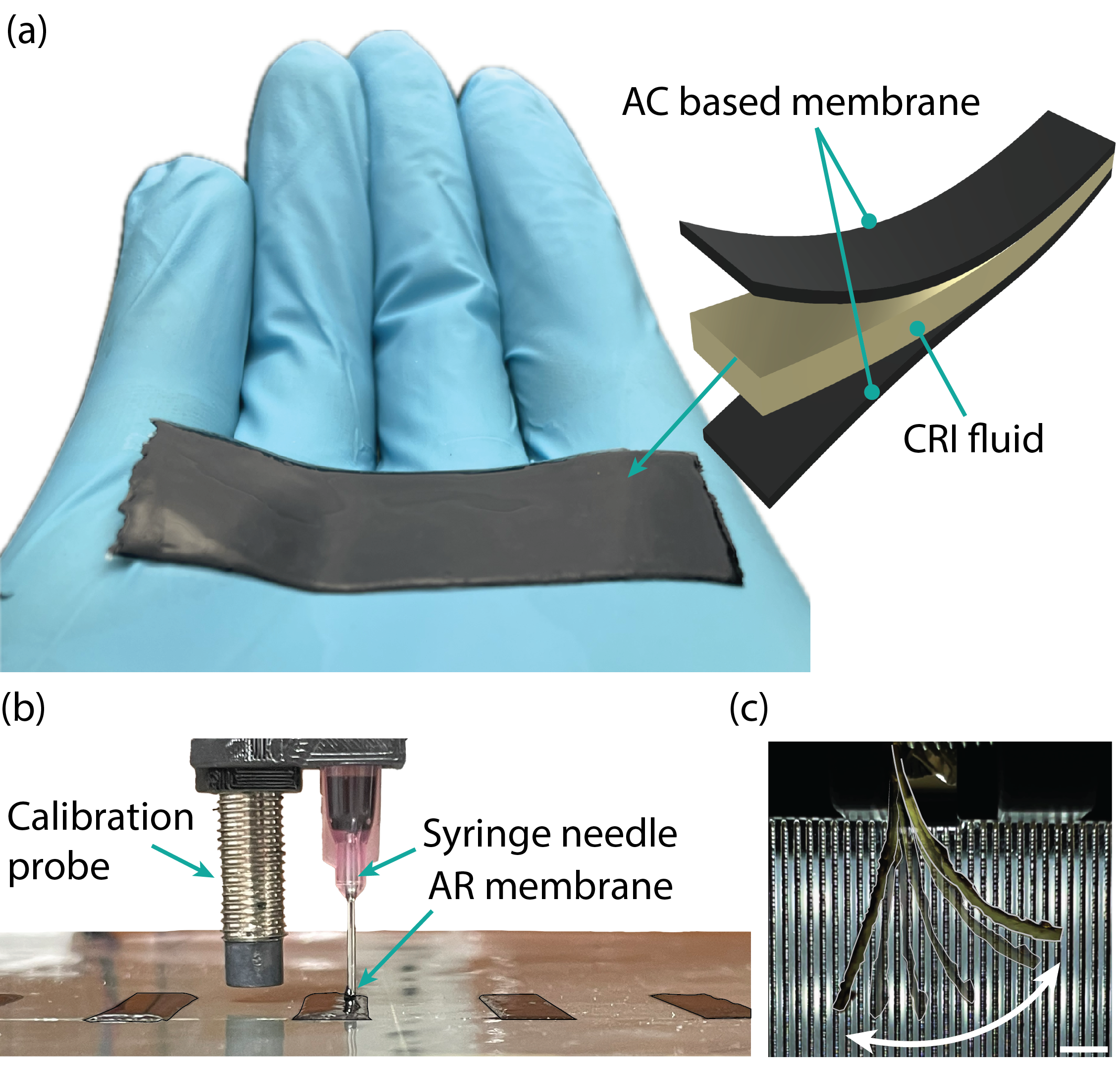}
\setlength{\belowcaptionskip}{-10pt}
\caption{a) Concept, structure, and development of ionic polymer actuator. b) Extruded material by DIW method. b) CRI layer developed by layer-by-layer process.
fabrication. c) Trajectory of bending motion of ionic polymer actuators. Scale bar is 5 mm}
\label{fig:fig1}
\end{figure}

Ionic polymer actuators have garnered considerable research interest owing to their fast response, lightweight, high flexibility, large bending strain, noiseless operation, and low voltage activation ($\leq$ 10 V) \cite{shahinpoor2003ionic}. The operation voltage is lower than the safety threshold for the human body, ensuring that ionic polymer actuators are safe around humans \cite{lu2023ionic}.

Typically ionic polymer actuators consist of an active ionic exchange polymer or fluid membrane sandwiched between electrodes. When voltage is applied, cations and anions within the ionic polymer migrate towards opposite electrodes, causing anisotropic polymer expansion that results in a bending motion \cite{zhang2023low}. To enhance this bending performance, researchers have explored various modifications to the polymer materials and electrode designs, such as Nafion\textsuperscript{TM}, polyvinylidene fluoride, and single or double-walled carbon nanotubes \cite{terasawa2014high, mukai2009highly}. However, these materials are toxic, and the membranes are brittle, posing significant risks when interfacing and interacting with humans. To address this issue, researchers have developed biocompatible ionic polymer actuators with materials such as chitosan, glycerol, and genipin \cite{cheedarala2019ionic, nevstrueva2018natural, jeon2013dry, romero2014biocompatible,  zhao2017development, he2015ionic, rajagopalan2011fullerenol, nan2020high}. However, fully biocompatible ionic polymer actuators are challenging since they still utilize Nafion\textsuperscript{TM}, sulfuric acid, or hydrogen chloride to bond ionic polymer membranes with electrodes or for compound treatment. Even in cases where full biocompatibility has been achieved, these ionic polymer actuators often suffer from reduced performance \cite{cheedarala2019ionic, romero2014biocompatible}. The main challenge lies in ensuring encapsulation between ionic polymer and electrode while maintaining an efficient ion pathway. 

In this work, we present ionic polymer actuators utilizing biocompatible polymers, as shown in Fig.~\ref{fig:fig1}a. It is developed by utilizing a cellulose-reinforced ionic (CRI) fluid, activated carbon (AC) based membrane as the substrate material, and gold foil as electrodes. The ionic fluid, 1-Ethyl-3-methylimidazolium acetate ([EMIM][OAc]), is a biodegradable material \cite{sun2009complete} with low toxicity (LD${_{50}}$: 2000 - 5000 mg$\cdot \text{kg}^{-1}$ (Rat) \cite{fisher_sci_url}). It exhibits a non-Newtonian fluid behavior, has a moderate viscosity of $\approx$ 140 mPa$\cdot$s \cite{quijada2012experimental,gomez2006physical}, and is thermally stable $\leq$ 100 $^\circ$C \cite{williams2018thermal}. Although the CRI fluid is not biocompatible, the biocompatible AC-based ink encapsulates it completely to form a biocompatible membrane. The AC-based membrane also contributes to increasing the stiffness of the membrane and it still allows the exchange of cations and anions between the electrodes with moderate electrical conductivity. 

Finally, in this work we tackled the manufacturing of ionic polymer actuators with a direct ink-writing (DIW) method, as seen in Fig.~\ref{fig:fig1}b. We developed a method to perform the printing process in a single step, printing both the fluid and fluid substrate that encapsulates the structure together. The materials are extruded by a syringe needle (Test Equipment Depot, 920050-TE, MA) calibrated utilizing the onboard induction-based probe found on the Prusa i3 MK3S (Prusa, Prague, Czech Republic). The printed polymer membranes are attached to gold foils and voltage is applied to create bending motion as shown in Fig.~\ref{fig:fig1}c.

\text{Key contributions of this study include:}
\begin{enumerate}
    \item Development of a biocompatible actuator design where ionic polymer layers encapsulate a non-biocompatible substrate, ensuring enhanced compatibility for biomedical applications.
    \item Advancement of ionic polymer membrane fabrication tailored for soft actuators, achieved through a single-step DIW process, simplifying production workflows.
    \item Achievement of actuators demonstrating the highest curvature reported for biocompatible ionic polymer actuators, showcasing superior performance.
\end{enumerate}


The reminder of this paper is organized as follows. Section~\ref{sec:act_design_fab} introduces the design, development, and fabrication of ionic polymer actuators. Section~\ref{sec:characterization} describes the various characterization tests done on the ionic polymer actuators. Finally, Section~\ref{sec:conclude} concludes this paper and discusses future work.
%

\begin{figure}[t]
\centering
\includegraphics[width=0.49\textwidth]{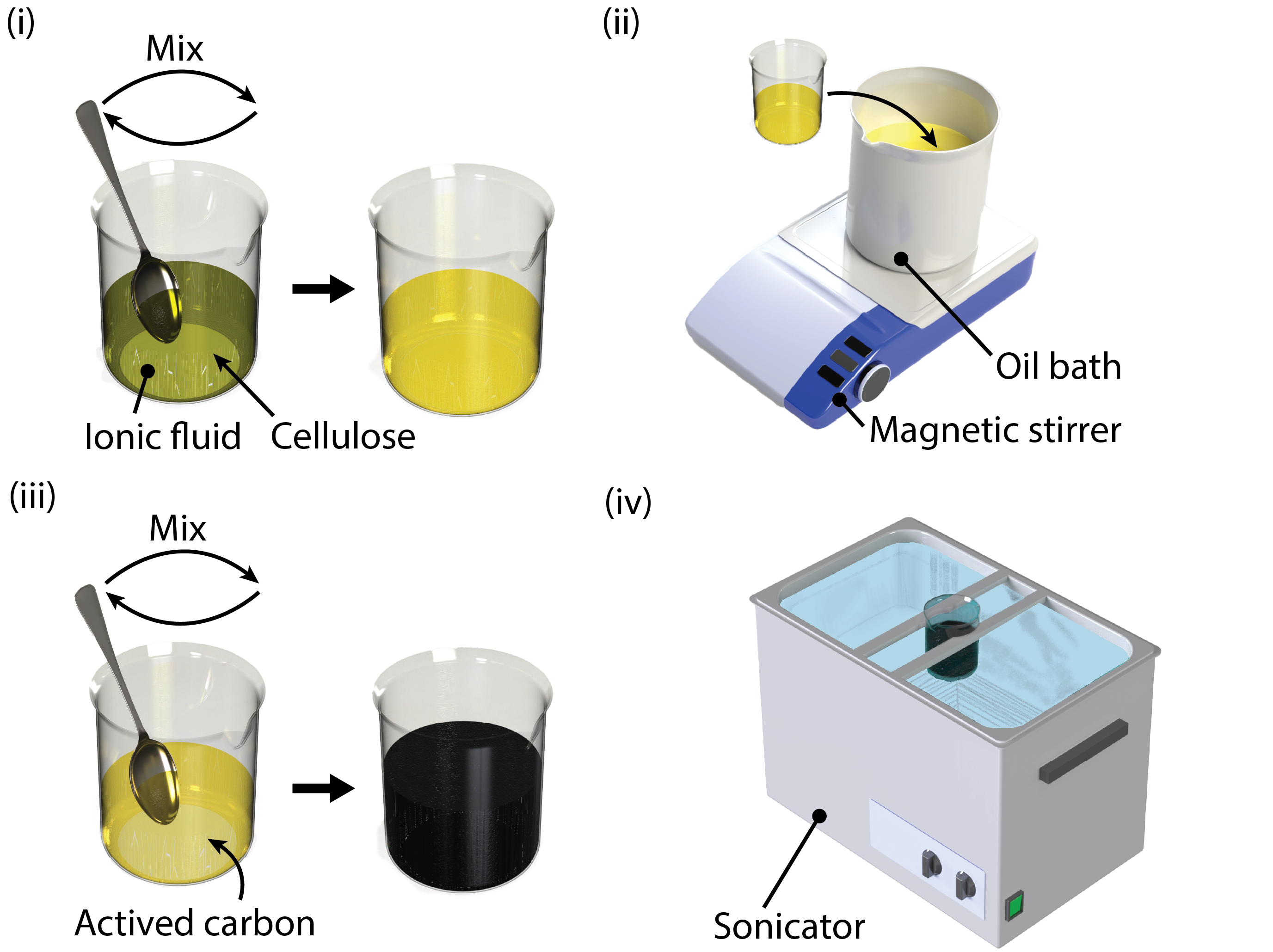}
\caption{Preparation scheme of ionic polymer inks. i) Cellulose mixed with ionic fluid. ii) Ionic fluid heated by oil bath. iii) Mixing AC. iv) Particle dispersion by sonicator.}
\label{fig:fabrication}
\end{figure}

\section{Developemnt of ionic polymer actuators}
\label{sec:act_design_fab}
%
\subsection{Ink preparation for DIW}
The polymer inks for the CRI and AC were prepared as highlighted in Fig.~\ref{fig:fabrication}. To create the CRI ink, the 9 wt\% of cellulose (9004-34-6, Carl Roth, Germany) was gradually mixed with [EMIM][OAc] (143314-17-4, Proionic, Austria) using a spoon and then placed in a 90 $^{\circ}$C oil bath under a magnetic stirrer (Sobocat HC, Südstärke GmbH, Germany) at 200 rpm for one hour. To create the AR-based ink, activated carbon (C4386, Sigma-Aldrich, MA) is added to a part of the previous mixture, with the following mixing ratio: 32 wt\% of activated carbon, 4 wt\% of cellulose, 64 wt\% of [EMIM][OAc]. The final mixture was placed in ultrasonic sonicator (VEVOR 30L, VEVOR, China) to disperse particles to obtain the AC-based ink. Both inks were added into a 1 mL capacity syringe. This small-capacity syringe helps to improve extruding resolution owing to the long Z-axis travel distance. 

\begin{figure}[t!]
\centering
\includegraphics[width=0.49\textwidth]{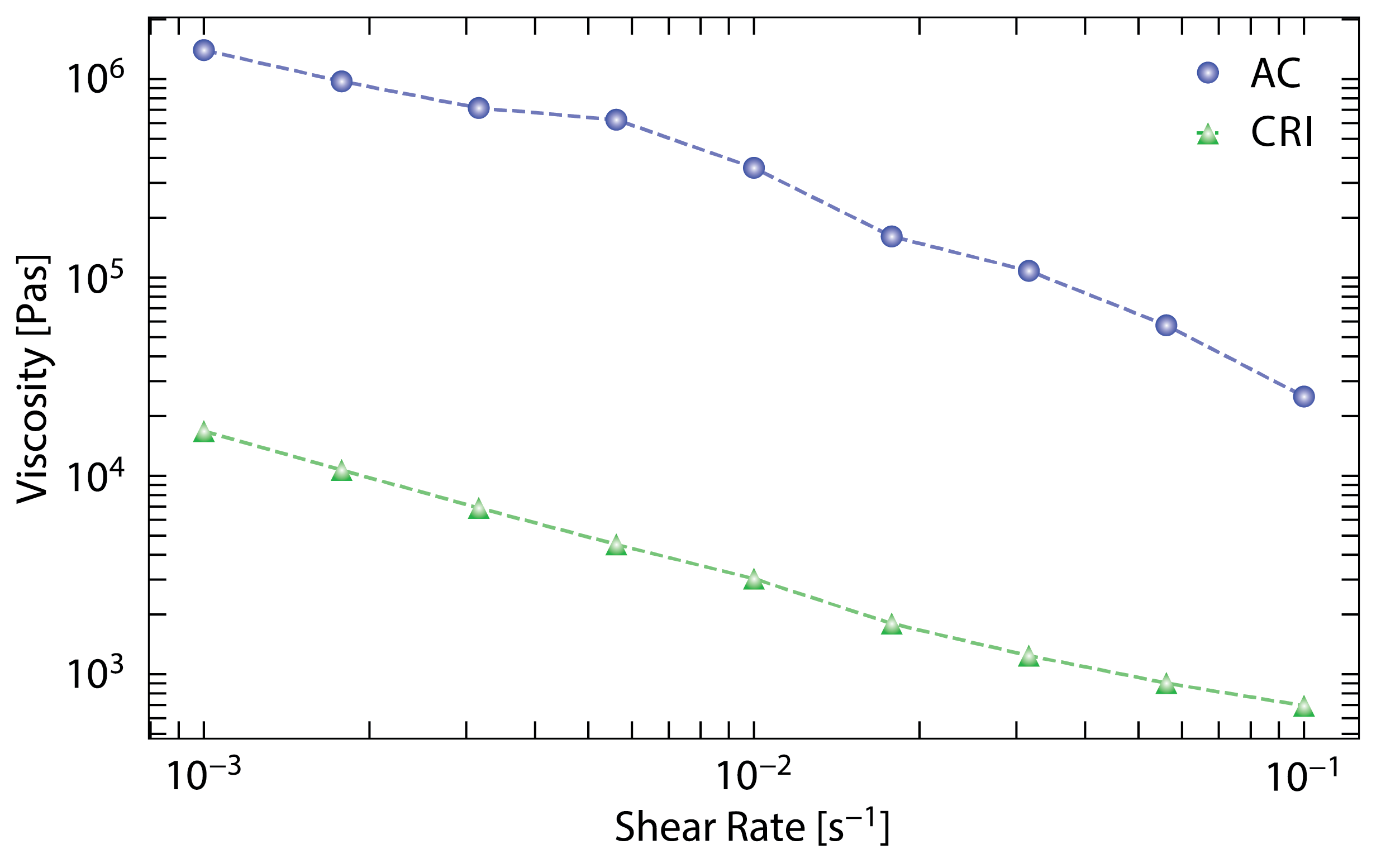}
\caption{Relationship between shear rate of CRI and AC fluids measured by rheometer.}
\label{fig:rheology}
\end{figure}


\begin{figure}[b]
\centering
\includegraphics[width=0.49\textwidth]{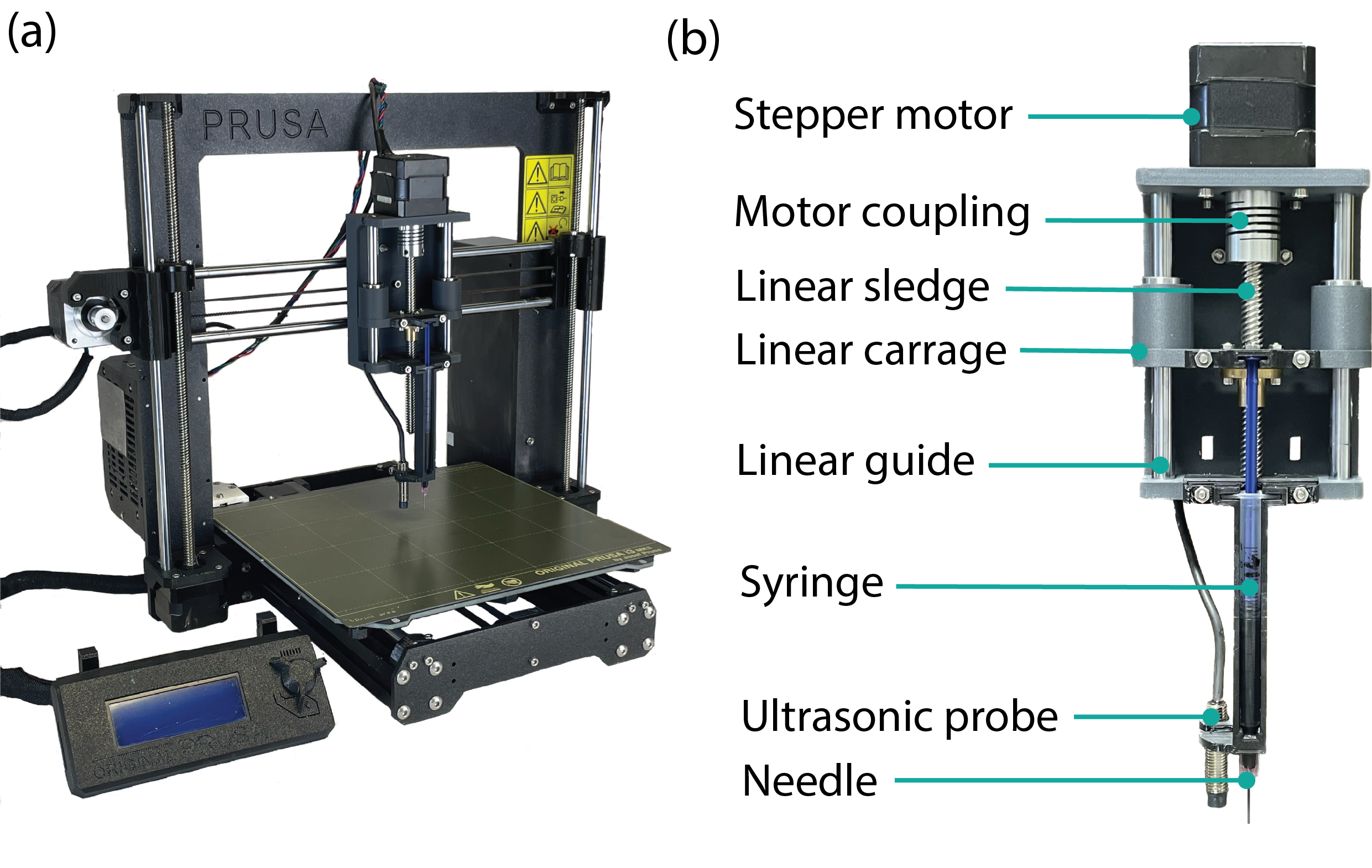}
\caption{Customized 3D printer. a) Printer overview. b) Ink extrusion system. It is installed at the place of the nozzle head of a 3d printer.}
\label{fig:DIW}
\end{figure}


\subsection{Characterization of ink}
The viscosities and rheological behavior of both inks was important to understand as we were extruding the inks at different rates while printing. Thus, we measured the viscosities by rheometer (Physica MCR301, Anton Paar, Austria), as shown in Fig.~\ref{fig:rheology}. What was noticed was that the inks would dry out over time and form lumps, thus these data were collected prior to any clump formation or drying affects. Based on the results, both fluids have shear-thinning behavior which is suitable for DIW. 

Since the AC-based ink requires specific properties to effectively encapsulate the CRI ink while providing sufficient conductivity for ion migration, we measured the resistance of this ink to evaluate this property utilizing a digital multimeter (DMM6500, Keithley, OH) on three samples. The dimensions of the samples were 10 mm in length and 10 mm in width. The thickness of every sample was measured using a laser displacement sensor (Sick, OD5000, Germany). Across samples, the thickness was 200 $\mu$m. The measured sheet resistance is 4.36 $\pm$ 0.21 k$\Omega \ \mathrm{sq}^{-1}$. Although this value is not high for conductive materials, it is still in the range of in line with semi-conductive materials.

In addition, the density of inks was also characterized. 1 mL pycnometers (12797, VWR International, PA) were utilized to fill CRI and AC-based inks. For each ink, three samples were measured. The measured density of CRI ink is 1.13 $\pm$ 0.07 g $\text{cm}^{-3}$, and AC-based ink is 1.23 $\pm$ 0.12 g $\text{cm}^{-3}$. Since both fluids are heavier than water, they would be able to operate both air and underwater.


\subsection{Extrusion system}
Prepared inks in syringe cartridges were extruded utilizing a customized low-cost 3D printer (Original Prusa i3 MK3S+, Prusa Research, Czech), as shown in Fig.~\ref{fig:DIW}a. From Fig.~\ref{fig:DIW}a and b, some of the customizations include the extruder and the firmware running the print. In hardware setup, the filament extrusion system was replaced with the new ink extrusion system, as seen in Fig.~\ref{fig:DIW}b. The system utilizes a SuperPINDA ultrasonic probe for calibration and the ink is extruded by a linear guided Prusa stepper motor. The linear sledge, motor coupling, linear guide, and syringe were purchased from distributors in Switzerland. The syringe and motor housing were 3D printed with PLA material (X1 Carbon, Bambu Lab EU, Germany). In the software, the firmware deployed is Marlin, which enables high customization. To execute the print functions, G-code commands were processed using FullControlXYZ, an open-source Python-based library. The generated G-code was uploaded and executed using the slicer software Repetier-Host (Hot-World GmbH \& Co. KG, Germany). The AC-based ink is printed first as the substrate material, after which the syringe containing the CRI ink is swapped in. Afterward, the printer returns to its original position to align and extrude any residual AC-based ink remaining in the syringe tip. Following this, CRI layers are printed. In the final step, the AC-based syringe is swapped back in and the top layer substrate is printed by positioning and extruding the remaining CRI ink. The completed membrane can be seen in Fig.~\ref{fig:fig1}a, with a width and length of 6 and 30 mm, subsequently. The entire printing process is shown in the supplementary video. In the video, 59 wt\% of extruded inks were wasted, and 41 wt\% inks were utilized for the membrane. However, the amount of waste inks can relatively be reduced to 20 wt\% if multiple batches are printed, with 80 wt\% of the ink being used for membranes. The printed membrane was immersed in deionized water for 24 h for coagulation. Finally, prior to testing the two sheets of gold foil were attached to the coagulated membrane, with [EMIM][OAc] utilized to increase the bonding strength between them.



\begin{figure}[t]
\centering
\setlength{\belowcaptionskip}{2pt}
\includegraphics[width=0.49\textwidth]{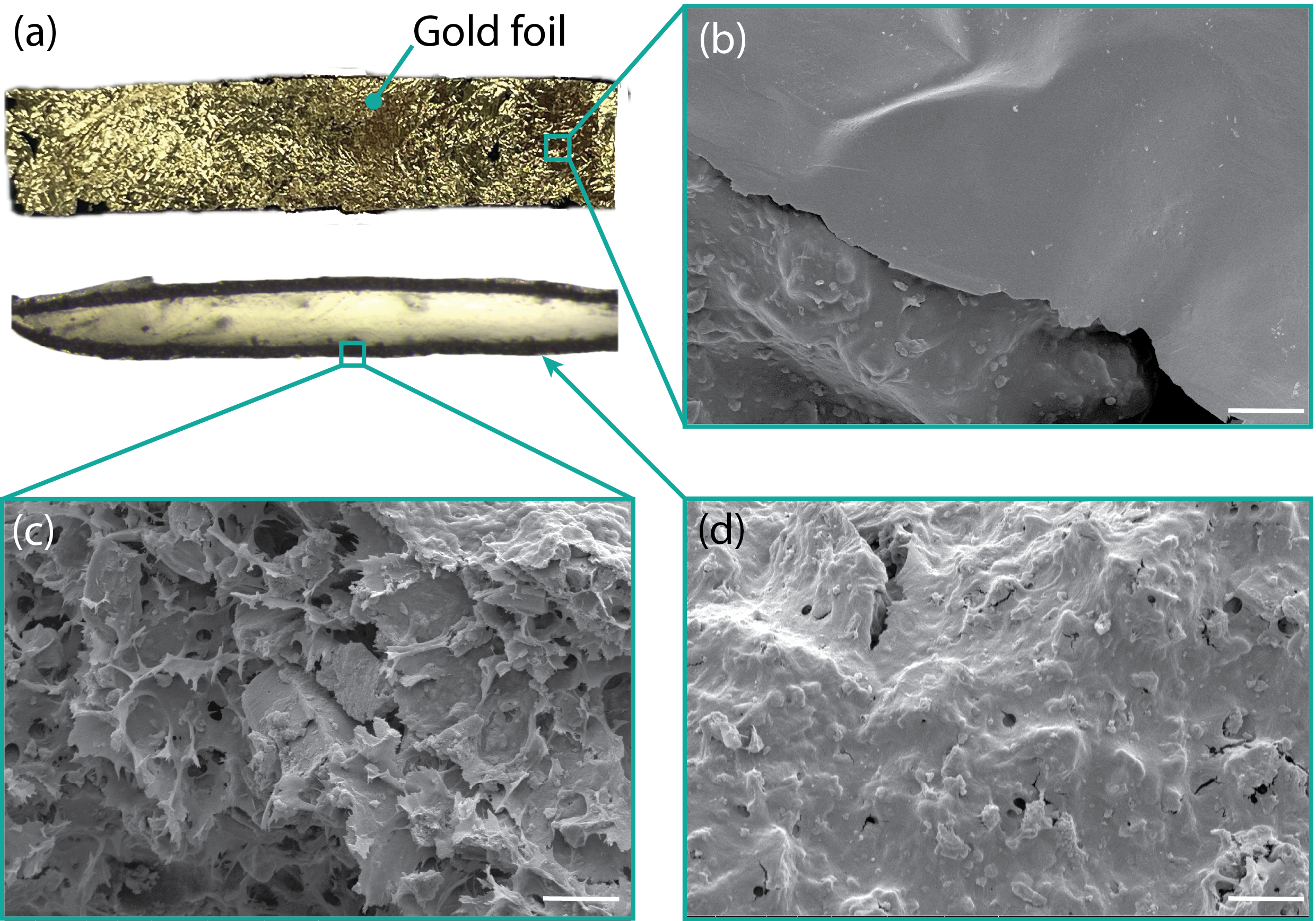}
\caption{Microscopic image of ionic polymer actuators. a) Front and cross-section view by optical microscope. b) SEM images of the front view of gold foil. c) Cross-section view of AC-based membrane. d) Front view of AC-based membrane. All the scale bars represent 5 $\mu$m.}
\label{fig:SEM_Image}
\end{figure}
\begin{figure}[b]
\centering
\includegraphics[width=0.49\textwidth]{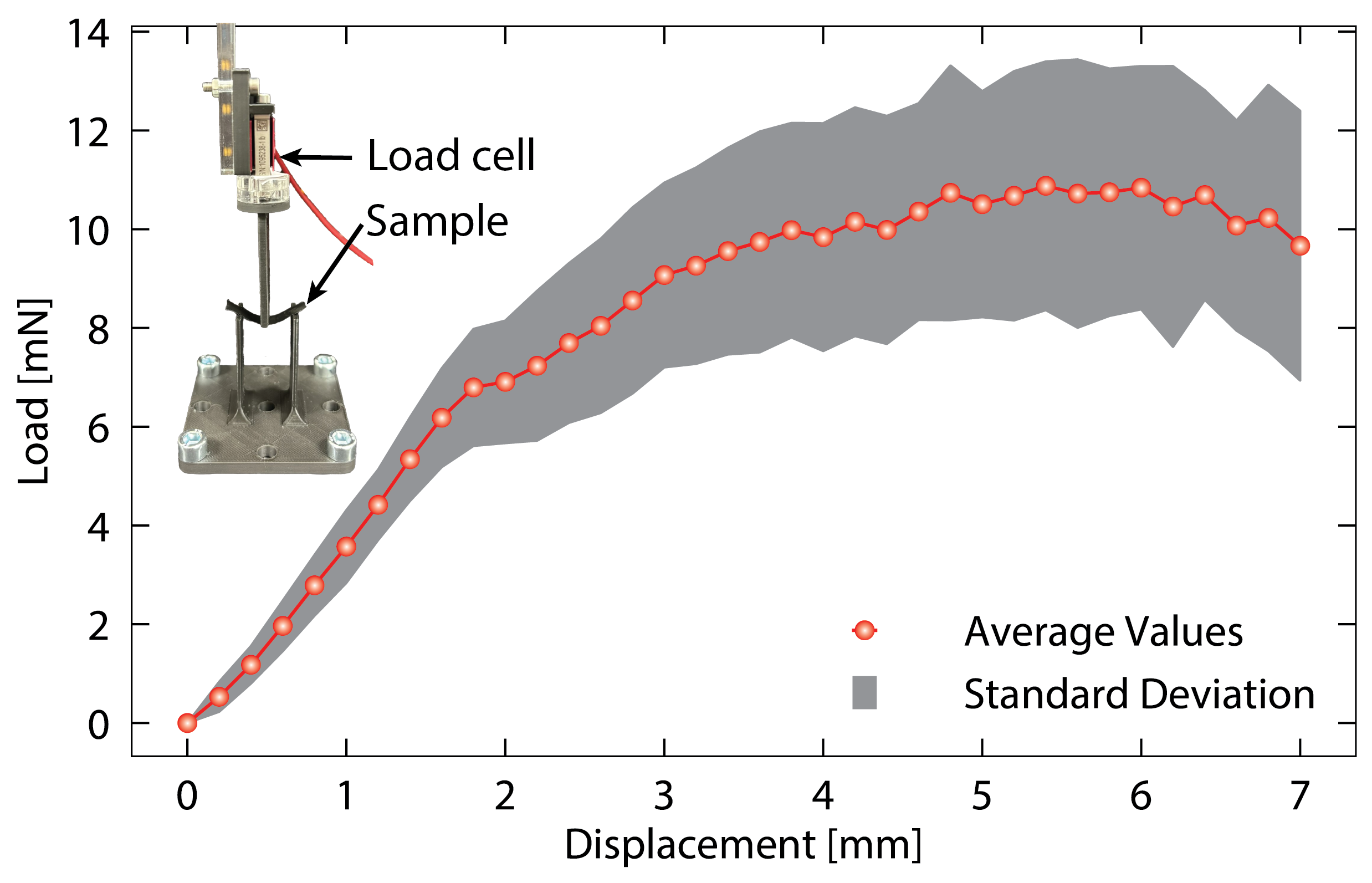}
\caption{Relationship between load of ionic polymer membrane (n = 3) and displacement of linear stage.}
\label{fig:3point_test}
\end{figure}
\section{Actuator characterization}
\label{sec:characterization}

\begin{figure}[]
\centering
\includegraphics[width=0.49\textwidth]{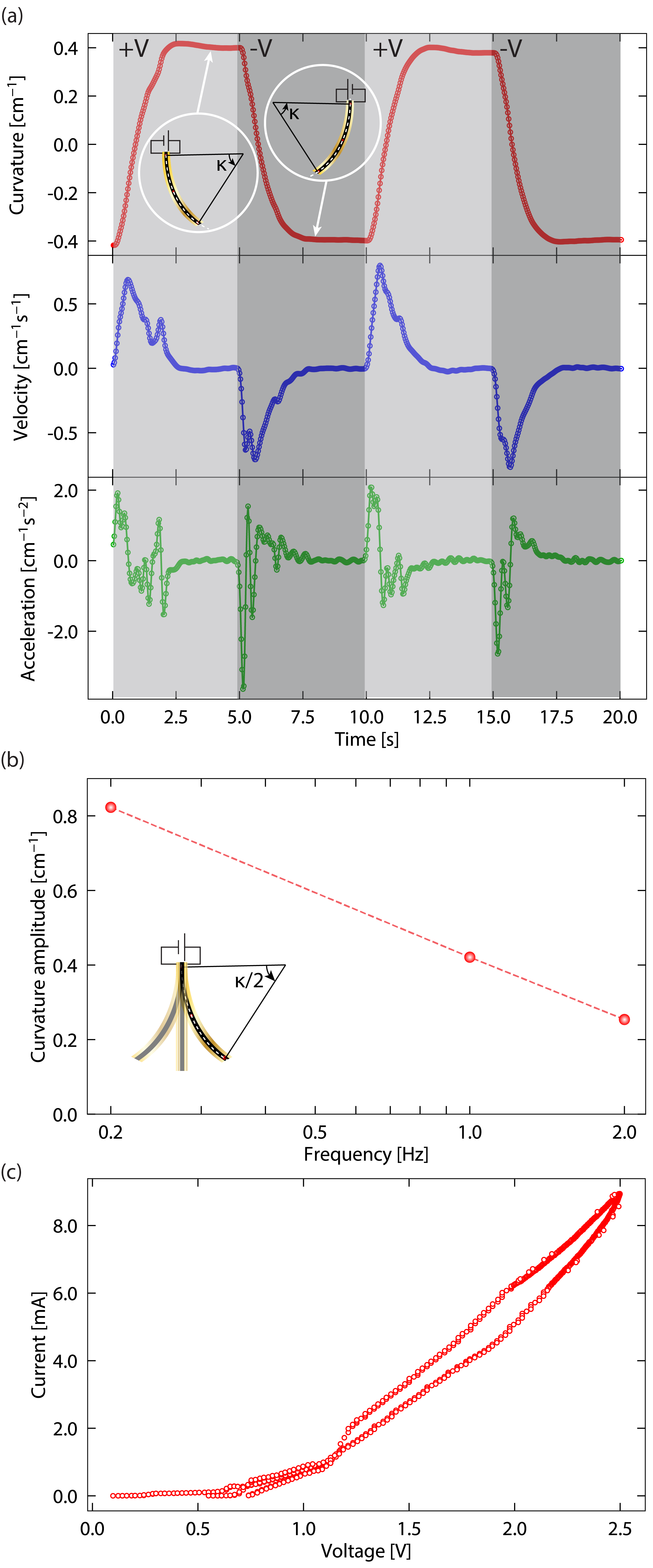}
\caption{Characterization of bending motion of ionic polymer actuators applying 4.5 V. a) Curvature, velocity, and acceleration versus time. b) Curvature amplitude versus frequency. c) Current versus Voltage.}
\label{fig:frequency_test}
\end{figure}

The developed ionic polymer actuators were examined using SEM and optical microscopy to assess the print quality, followed by comprehensive mechanical, electrical, and performance testing. All actuator characterizations were performed under atmospheric conditions.

\subsection{Microscopic observation}
The cross-sectional view shown in  Fig.~\ref{fig:SEM_Image} was taken with an optical microscope (Stemi 508, Zeiss, Germany), highlighting the uniform layers of the AC and CRI fluid. Further detailed imaging was taken to inspect the layers utilizing a Scanning Electron Microscope, SEM (Quanta 650 FEG ESEM, FEI, ORE), as seen in Figs.~\ref{fig:SEM_Image}b - d. To prepare the samples, they were glued to SEM aluminum holders with a double-sided carbon pad and coated with a 7 nm thin platinum layer on a sputter coater (MED020, Bal-Tec, Liechtenstein). SEM was performed at an acceleration voltage of 5 kV and a working distance of 10 mm. Figure~\ref{fig:SEM_Image}b shows that the gold foil is deposited as a layer on the printed AC. From Fig.~\ref{fig:SEM_Image}c, we notice the fractured cross-section of the layers, highlighting a relatively uniform porous structure in the AC-based ink due to the addition of the cellulose. We noticed some larger agglomerations of carbon as well. In Fig.~\ref{fig:SEM_Image}d, we noticed the surface of the printed AC which was more rough and relatively less porous.

\subsection{Three-point bending test}
To characterize the stiffness of the IAP actuators, we performed a three-point bending test because the samples were a bit too delicate for the uniaxial tensile testing machine. The test setup was developed to match the size of each actuator, as seen in Fig.~\ref{fig:3point_test}. The probe was connected to a miniature load cell (FSH04085, Futek, CA) and was adjusted to travel linearly utilizing a 1 $\mu$m resolution linear stage (PT1/M, Thorlabs, NJ) at increments of 200 $\mu$m. From the results in Fig.~\ref{fig:3point_test}, a linear deformation is observed up to approximately 1.6 mm of displacement. Beyond this point, the behavior transitions to non-linear until reaching 7 mm, where plastic deformation appears to begin. Within the range of linear trend, Young's modulus $E$ of ionic polymer actuators can be calculated by (\ref{eq:elastic_modulus}), where $L$ is the span length between supports, $b$ is the width of samples (6 mm), $d$ is the thickness of samples (0.67 mm). The terms $\Delta F$ and $\Delta \delta$ are loading value and displacement, respectively.

\begin{equation}
E = \frac{L^3}{4bd^3} \times \frac{\Delta F}{\Delta \delta}
\label{eq:elastic_modulus}
\end{equation}

$\Delta F / \Delta \delta$ was calculated utilizing linear fit from 0 to 1.6 mm displacement. The $E$ value is calculated as 0.40 $\pm$ 0.06 MPa, which indicates that the material exhibits a low stiffness, making it suitable for use as ionic polymer actuators, similar to previous studies \cite{wang2017high}.

\subsection{Curvature characterization}
The performance of the ionic polymer actuators was characterized by applying an input voltage between 2.5 and 4.5 V, which induced a bending motion. Beyond the input voltage 4.5 V, the actuators would fail. The voltage was supplied utilizing a DC power supply (PPS-16005, Voltcraft, Germany) and regulated by a microcontroller (Arduino UNO, Arduino, Italy) connected to a 4-channel relay module (WPM400, Velleman, Belgium). Actuators were fixed vertically, parallel to the force of gravity, and bent upwards, as shown in Fig.~\ref{fig:frequency_test} and in the supplementary video captured by APS-C camera (Z50, Nikon, Japan). The curvature performance of the actuators was analyzed utilizing trajectory tracking software (Tracker, Douglas Brown). The tracking data was processed by Gaussian filter ($\sigma$ = 2), as is highlighted in Fig.~\ref{fig:frequency_test}. From Fig.~\ref{fig:frequency_test}a, we notice the change in curvature, velocity, and acceleration over time. The maximum curvature produced by the actuator was 0.82 $\text{cm}^{-1}$. We hypothesize that this maximum curvature angle is limited by the internal mechanical resistance that counteracts the maximum possible bending angle. The curvature velocity reached its maximum amplitude at around 0.5 s, followed by a decrease until 2.5s. And the maximum velocity amplitude is 0.80 $\text{cm}^{-1}$$\text{s}^{-1}$. From the acceleration data, we note that it reaches its maximum acceleration of 3.64 $\text{cm}^{-1}$$\text{s}^{-2}$, upon voltage input. 

Additionally, strain can also be calculated using (\ref{eq:strain}):

\begin{equation}
\epsilon_{\text{max}} = \frac{2d\delta_{\text{max}}}{L^2 + {\delta_{\text{max}}}^2}
\label{eq:strain}
\end{equation}

where $\delta_{\text{max}}$ is the maximum tip displacement \cite{he2015ionic}. Since the value of $\delta_{\text{max}}$ was measured to be 11.61 mm, the maximum strain is calculated to be 1.50 \%. 

At different applied frequencies of 0.2, 1, and 2 Hz, the curvature amplitude was measured to be 0.82 $\text{cm}^{-1}$, 0.42 $\text{cm}^{-1}$, 0.25 $\text{cm}^{-1}$, as highlighted in Fig.~\ref{fig:frequency_test}b. After 2 Hz, the actuator vibrates with almost no bending changes, following the trend with increasing frequency as seen in Fig.~\ref{fig:frequency_test}b. For the consumption power of actuators, current and voltage was measured during actuation. The maximum value was 8.9 mA at 2.5 V.

\begin{figure}[t]
\centering
\setlength{\belowcaptionskip}{2pt}
\includegraphics[width=0.49\textwidth]{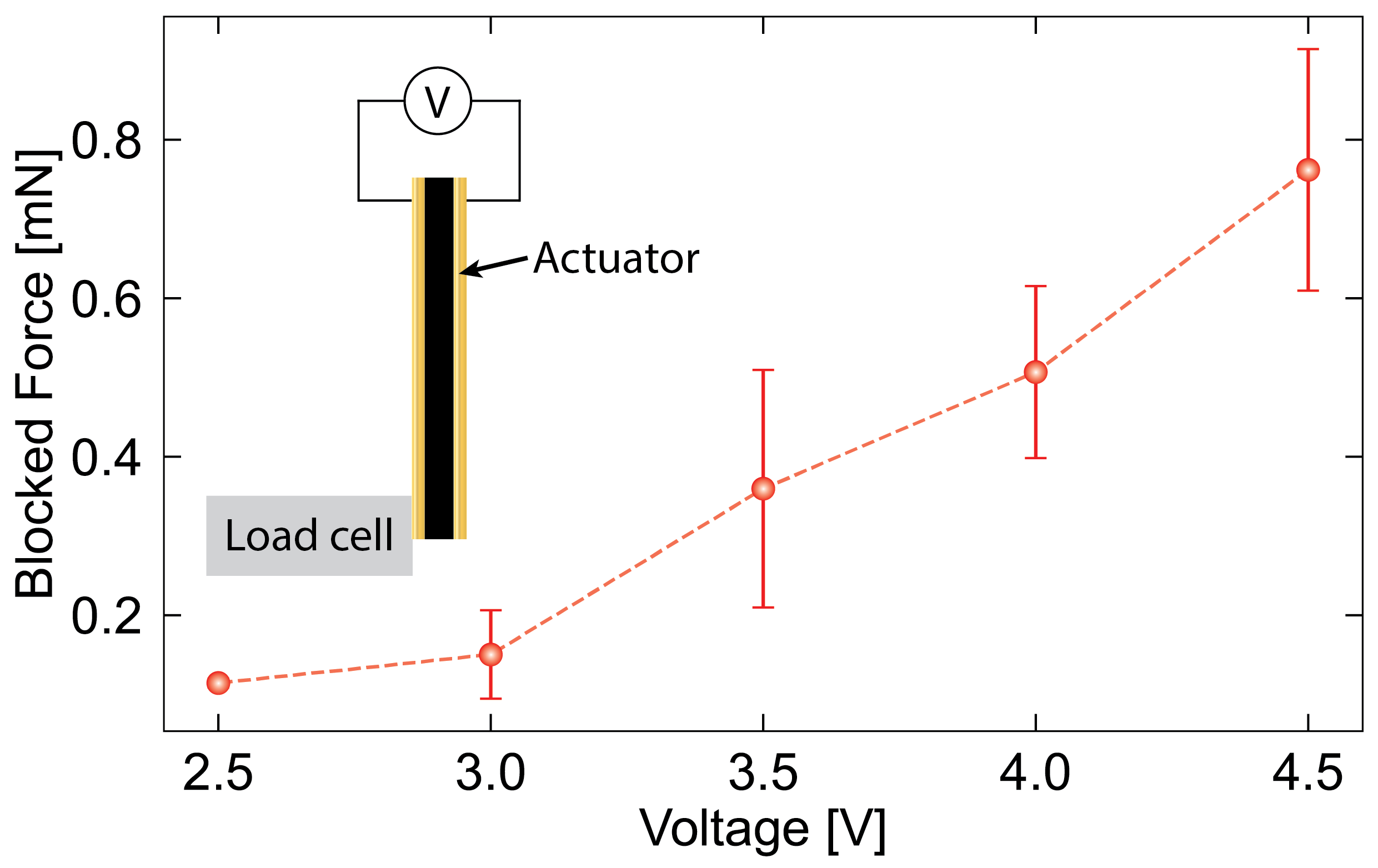}
\caption{Blocked force versus voltage of ionic polymer actuators (n = 3). The tip of actuators is attached to the load cell.}
\label{fig:bending_force}
\end{figure}

\begin{table*}[t]
\centering
\caption{Performance comparison}
\label{tbl:performance_comparison}
\resizebox{0.95\textwidth}{!}{%
\begin{tabular}{llccc}
\toprule
\text{Reference}                      & \text{Base material} & \text{Max curvature} [$\text{cm}^{-1}$] & \text{Max strain [\%]} & \text{Blocked force [mN]} \\ 
\midrule
\text{This work}                      & Activated carbon, Cellulose, [EMIM][OAc]                     & 0.82                                   & 1.50                     & 0.76                        \\ 
\cite{cheedarala2019ionic}              & Chitosan, Graphene oxide, [EMIM][N(Tf)$_{2}$]                     & N/A                                    & $\approx$ 0.09           & N/A                         \\ 
\cite{nevstrueva2018natural}            & Nafion\textsuperscript{TM}, Cellulose, [EMIm][OAc]                 & 0.8                                    & $\approx$ 0.6            & N/A                         \\ 
\cite{jeon2013dry}                      & Chitosan, Graphene oxide, [EMIM][OTf]                 & 0.05                                    & N/A                     & N/A                         \\ 
\cite{romero2014biocompatible}          & Silk fibroin, Polypyrrole                 & 0.05                                    & N/A                     & $\approx$ 10                \\ 
\cite{zhao2017development}              & Chitosan, MCNT, BMImBF$_{4}$                 & N/A                                    & $\approx$ 0.77           & $\approx$ 1.6               \\ 
\cite{he2015ionic}                      & Chitosan, Graphene, BMImBF$_{4}$                 & N/A                                    & 0.1                      & N/A                         \\ 
\cite{rajagopalan2011fullerenol}        & PHF, SPEI                 & N/A                                    & 0.05                     & 6.60                        \\ 
\cite{nan2020high}                      & PVDF, Cellulose Acetate, Graphene                 & N/A                                    & $\approx$ 0.15           & 0.78                        \\ 
\bottomrule
\end{tabular}%
}
\end{table*}



\subsection{Blocked force characterization}
To measure the blocked force of the actuators, the actuator was aligned against a load cell (PCB 250-3, KERN, Germany) as seen in Fig.~\ref{fig:bending_force}. The applied voltage to the actuator was varied from 2.5 V to 4.5 V in increments of 0.5 V, and measurements were averaged 5 s after the voltage input. The results highlighted increasing blocked force reading with the increasing applied voltage. The maximum force was measured to be 0.76 $\pm$ 0.15 mN, as seen in Fig.~\ref{fig:bending_force}. 

According to Table.~\ref{tbl:performance_comparison} that highlights the performance of previous work and ours, the actuators in this work achieve the highest curvature and strain values with comparable blocked force. The force could be further increased by simply scaling up the actuator dimensions. Velocities of the actuators were not included in the table because it was not a consistent metric we could obtain from previous work, but our actuator demonstrated a relative velocity of 0.80 $\text{cm}^{-1}$$\text{s}^{-1}$.

\section{Conclusion and Future Work}
\label{sec:conclude}

In this work, we presented a biocompatible, low-cost, fully 3D-printed IAP membrane actuator. The membrane consisted of layers of activated carbon (AC) and a cellulose-reinforced ionic (CRI). Due to the CRI layer that encapsulates the AC, the IAP actuator exhibited biocompatibility properties. The actuator demonstrated promising bending performance. By utilizing a low-cost 3D printer, we were able to achieve a one-step direct ink writing process that prints both the CRI fluid and AC-based membrane substrate, offering the community a straightforward manufacturing method. Our results showed that the actuator is capable of achieving a maximum bending curvature of 0.82 $\text{cm}^{-1}$, which outperforms many of the existing ionic polymer actuators. It also provides a blocked force of up to 0.76 mN matching what is seen in literature. We believe that the combination between its performance, flexibility, and low-voltage requirements will make it a suitable candidate for a wide range of soft robotic applications in the air and water. 

Future research could extend to numerous efforts, in the fields of environmental monitoring and human-interactive applications. Since these actuators do not dry out underwater, we seem the utilization of these actuators for aquatic tasks in the future as well. However, an insulating coating might be needed to prevent short circuits that could disable or damage the actuators. The low driving voltage, under 10 V, ensures that the actuators are safe and that the driving electronics can be designed for untethered applications. Further, because of the manufacturing setup we develop, we aim to explore further customization of these actuators, for different functional requirements. 

To improve the actuator's outdoor performance, we are looking into optimizing the material properties of the actuator for improved blocked force, increasing operational lifespan, and overall durability, with the aim of maintaining its biocompatibility. The second goal would be to expand its function away from just actuation and test it for its sensing capabilities as well. Since the structure of actuators has two electrodes sandwiching ionic polymer, actuators could work as capacitors with self-sensing. Thus, we envision the developed actuators with self-sensing capabilities for future soft robotic systems. 

In conclusion, our biocompatible, fully 3D-printed IAP actuator, represents advancement in the field of biocompatible ionic polymer membrane actuators. Its combination of low-cost, one-shot fabrication and promising high-frequency, high bending, and blocked forced characteristics makes it a strong candidate for a variety of applications that require an actuator that can interact with more fragile and sensitive systems. We believe this is just a first step towards a more transient high-performing soft actuator that can support sustainable approaches to building soft robots in the future.


\section*{ACKNOWLEDGMENTS}
We are deeply grateful to Dr. Gilberto Siqueira for his great assistance in characterizing rheology of ink.
We sincerely thank Dr. Ronan Hinchet for his remarks and suggestions throughout this work. 
We also thank the members of the Laboratory of
Sustainability Robotics for their
support and stimulating discussions on this topic.
This work was supported in part by EPSRC Awards (grant no, EP/R009953/1), in part by the Swiss National Science Foundation (SNSF) International Cooperation, EU ERANet grant (grant no, 194986), and in part by the ERC Consolidator Grant as funded by (State Secretariat for Education, Research, and Innovation) SERI (grant no. MB22.00066). 

\end{document}